\documentclass{article}

\usepackage[T1]{fontenc}
\usepackage{lmodern}

\usepackage[utf8]{inputenc} 
\usepackage[T1]{fontenc}    
\usepackage{hyperref}       
\usepackage{url}            
\usepackage{booktabs}       
\usepackage{amsfonts}       
\usepackage{nicefrac}       
\usepackage{microtype}      
\usepackage{color}	    
 \usepackage{amssymb,amsmath} 
\usepackage{epsfig,graphicx}

\usepackage{caption}
\usepackage{subcaption}

 \def\R{\mathbb{R}}
 \def\D{\mathcal{D}}
 \def\d{\mathrm{d}}
 \def\dt{\mathrm{d}t}
 \def\dx{\mathrm{d}x}
 \def\div{\mathrm{div}} 
 \def\E{\mathrm{E}}
 
 \def\cov{\mathrm{cov}} 
 \def\wcov{\mathrm{wcov}} 
 \def\min{\mathrm{min}} 
 \def\max{\mathrm{max}} 
 \def\deltat{{\delta t}} 
 \def\h{\deltat} 
 \def\rR{\mathrm{R}}

\def\Nabla{\nabla}
\title{Ensemble preconditioning for Markov chain Monte Carlo simulation}

\author{
  Charles Matthews\thanks{corresponding author} \\
  Department of Statistics \\
  University of Chicago\\
  Chicago IL, 60615 \\
  \texttt{c.matthews@uchicago.edu} \\
   \and
   Jonathan Weare \\
  Department of Statistics and\\
  James Frank Institute\\
  University of Chicago\\
  Chicago IL, 60615 \\
  \texttt{weare@uchicago.edu} \\
   \and
   Benedict Leimkuhler \\
   School of Mathematics \\
   University of Edinburgh \\
   Edinburgh, EH93FD, UK\\
   \texttt{b.leimkuhler@ed.ac.uk} \\
}

\begin{document}

\maketitle

\begin{abstract}
    We describe parallel Markov chain Monte Carlo methods that propagate a collective ensemble of paths, with local covariance information calculated from neighboring replicas.   The use of collective dynamics eliminates multiplicative noise and stabilizes the dynamics thus providing a practical approach to difficult anisotropic sampling problems in high dimensions. Numerical experiments with model problems demonstrate that dramatic potential speedups, compared to various alternative schemes, are attainable. 
\end{abstract}

\section{Introduction}

A popular family of methods for Bayesian parameterization in data analytics are derived as Markov chain Monte Carlo (MCMC) methods, 
including Hamiltonian (or hybrid) Monte Carlo (HMC)\cite{Duane87}, 
or the Metropolis adjusted Langevin algorithm (MALA)\cite{RoDoFr78}. 
 These methods involve proposals that are based on approximating a continuous time (stochastic) dynamics that exactly preserves the target (posterior) density, followed  by an accept/reject step to correct for approximation errors. 
In some cases the Metropolis--Hastings correction may be omitted, provided the consequent approximation bias is well controlled.
Efficient parameterization of the stochastic differential equations used in these procedures has the potential to greatly accelerate their convergence, particularly when the target density is poorly scaled, i.e., when the Hessian matrix of the logarithm of the density has a large condition number.    
In precise analogy with well established strategies in optimization, the solution to conditioning problems in the sampling context is to find a well chosen change of variables (preconditioning) of the system, such that the natural scales of the transformed system are roughly commensurate.  


In this article we discuss an approach to dynamic preconditioning based on simultaneously evolving an ensemble of parallel MCMC simulations, each of which is referred to as a ``walker.''  As we will show, the walkers provide information that can greatly improve the efficiency of MCMC methods.   Our work is motivated by the method in \cite{GoWe} and earlier  similar schemes that use multiple parallel simulations to improve proposal densities in MCMC simulations (see e.g. \cite{GiRoGe1994,braak2006markov,christen2010general}).  The authors of \cite{GoWe} highlight the notion of affine invariance which means, loosely, that the performance of the method when applied to sample the probability distribution with density $\pi$ is identical to its performance when applied to sample  $\pi_{A,v}(x) \propto \pi(Ax+v)$ regardless of the choice of invertible matrix $A$ and vector $v.$   Affine invariant methods are advantageous when sampling a severely anisotropic density, or when a suitable parameterization length scale for the problem is unknown.  In this article, we however intentionally deviate from affine invariance to overcome practical limitations of earlier schemes.  The methods in this paper also differ from those in \cite{GoWe} in that, like the methods listed in the opening paragraph, they require the first derivative of $\pi$, giving high accuracy even in the absence of a Metropolis--Hastings step, and making the methods scalable to high dimension.


Our starting point is the discrete approximation of a system of stochastic differential equations (SDEs),  
\begin{equation}
\dot{x} = [J(x)+S(x)]\nabla \log\pi (x) + \div( J(x) + S(x) ) + \sqrt{ 2S(x)}\, \eta(t) \label{eqn::general}
\end{equation}
where $J(x)$ and $S(x)$ are skew-symmetric and symmetric positive semi-definite $N\times N$ matrices, respectively, with  $
\eta(t)$ representing a vector of independent Gaussian white noise components.   
In our sampling schemes, each walker generates a discrete time approximation of \eqref{eqn::general} with its own particular choice of $J$ which corresponds to a notion of the localized and regularized sample covariance matrix across the ensemble of walkers and incorporates information about the target density $\pi$ into the evolution of each walker.

Many optimization  methods can be characterized as time discretizations of \eqref{eqn::general}, with $J=0$ and choices of $S$ that encapsulate scaling information based upon higher derivatives of the model or a history of previously computed points.  Modification of $S$ to improve convergence also has been considered in the Monte Carlo literature, dating at least to \cite{bennett1975mass}.  This idea has been the focus of renewed attention in statistics and several new schemes of this or closely related types have been proposed \cite{Ghattas, GiCa11}.  From a more theoretical point of view, 
several authors have recently considered the effects of the choice of $J$ and $S$ on the ergodic properties of the solution \eqref{eqn::general} (see for example \cite{rey2015irreversible,DuLePa16,hwang2005accelerating,hwang1993accelerating}).  In this paper we are concerned with motivating and presenting a particular choice of $S$ and $J$ that yields practical and efficient sampling schemes and we will leave a more detailed analysis to later work.  Nonetheless,  our development  reveals the central role of discretization stability properties  in determining the efficacy of a particular choice of $J$ or $S.$  
\section{Preconditioning strategies for sampling \label{sec::preconditioning}}
As in any MCMC scheme, the goal is to estimate the average $\E[f] = \int\! f(x) \pi(x)\dx$  by a trajectory average of the form
 \[
\overline{f}_N =  \frac{1}{N}\sum_{n=0}^{n-1} f(x^{(n)}),
 \]
 for large $N$.  In many cases we can expect the error in an MCMC scheme to satisfy a central limit theorem:
 $
 \sqrt{N}\left( \overline{f}_N - \E[f] \right) \rightarrow_{\rm dist} N(0, \tau \sigma^2),
 $
where $\sigma^2$ is the variance of $f$ under $\pi$ (and is independent of the particular MCMC scheme), the quantity $\tau$ being the integrated auto-correlation time (IAT) which is often used to quantify the efficiency of an MCMC approach (see Appendix \ref{app::iat}). 
 
To emphasize an analogy with optimization, for the moment assume that $J=0.$  The steepest descent algorithm of optimization corresponds to the so-called overdamped Langevin (or Brownian) dynamics,
 \begin{equation}
 x^{(n+1)} = x^{(n)} + \deltat\, \Nabla\log(\pi(x^{(n)})) + \sqrt{2 \deltat}\, \rR^{(n)} \label{eqn::ovdld}
\end{equation}
where here we have used an Euler-Maruyama discretization \cite{TrMi04} and $\rR\sim N(0,I).$ Discretization introduces an $O(\deltat)$ error in the sampled invariant distribution so a Metropolis--Hastings accept/reject test may be incorporated in order to recover the correct statistics (see the MALA algorithm \cite{RoDoFr78}) when time discretization error dominates sampling error.  Reducing $\deltat$ gives a more accurate approximation of the evolution of the dynamics, and boosts the acceptance rate.   

When $\pi$ is Gaussian with covariance $\Sigma,$ one can easily show that the cost to achieve a fixed accuracy depends  on the condition number $\kappa = \lambda_{max}/\lambda_{min}$ where $\lambda_{max}$ and $\lambda_{min}$ are  the largest and smallest eigenvalues of $\Sigma.$  Indeed, one finds that the worst case IAT for the scheme in \eqref{eqn::ovdld} over observables of the form $v^\text{\tiny T} x$  is $\tau = \kappa -1$ (see Appendix \ref{app::iat}).  In this formula, the eigenvalue $\lambda_{min}$ arises due to the discretization stability constraint on the step-size parameter $\deltat$ and $\lambda_{max}$ appears because the direction of the corresponding eigenvector is slowest to relax for the continuous time process.  The presence of $\lambda_{min}$ in this formula indicates that analysis  of the continuous time scheme \eqref{eqn::general} (i.e. neglect of the discretization stability constraint)  can be misleading when considering the effects of poor conditioning on sampling efficiency.
Since the central limit theorem suggests that the error after $N$ steps of the scheme is roughly proportional to $ \sqrt{\tau/N},$ the cost to achieve a fixed accuracy is again roughly proportional to $\kappa.$

Continuing to use $J=0$ and taking, for example,  $S(x)=-(\nabla^2 \log(\pi(x)) )^{-1}$ in \eqref{eqn::general} and 
discretizing, we obtain a stochastic analogue of Newton's method,
\begin{equation}
x^{(n+1)} = x^{(n)} + \deltat\,S(x^{(n)}) \Nabla\log(\pi(x^{(n)}))  + \deltat\,\div S(x^{(n)}) + S^{1/2}(x^{(n)})  \sqrt{2 \deltat}\, \rR^{(n)}.
 \label{eqn::snewton}
\end{equation} 
Schemes of similar form though neglecting the $\div S$ term (and therefore requiring Metropolization) have been explored recently in \cite{Ghattas}.
Metropolization may also be used to correct the $\mathcal{O}(\deltat)$ sampling bias  introduced by the discretization.  It can be shown that the scheme is affine invariant in the sense that when it is applied to sampling $\pi_{A,v}$ it generates a sequence of samples $y^{(n)}$ so that $x^{(n)} = Ay^{(n)} +v$ has exactly the same distribution as the sequence of samples generated by the method when applied to $\pi.$   We therefore expect that when this method can be applied (e.g. when the Hessian is positive definite), it should be effective on poorly scaled problems.  This affine invariance property is shared by the deterministic Newton's method (obtained from \eqref{eqn::snewton} by dropping the noise term) and is responsible for its good performance when applied to optimize poorly scaled functions (e.g. when the condition number of the Hessian is large).  We stress that the key to the usefulness of either the deterministic or stochastic Newton's method is that one does not need to make an explicit choice of the matrix $A$ or the vector $v.$  
As the performance is independent of the choice of $A$ and $v$, we can assume that $A$ or $v$ are chosen to remove poor scaling within the problem. 

Due to the presence of the divergence term in the continuous dynamics, discretization will require evaluation of first, second and third order derivatives of $\log(\pi(x))$, making it prohibitively expensive for many models.   To avoid this difficulty, one can estimate the divergence term using an extra evaluation of the Hessian (see Appendix \ref{app::noisyestimation}), or omit the divergence term and rely on a Metropolization step to ensure correct sampling.
Regardless of how this term is handled,  the system  \eqref{eqn::snewton},  unlike \eqref{eqn::ovdld}, is based on multiplicative noise which is known to introduce complexity (and reduce accuracy) in numerical discretization \cite{TrMi04}.

\begin{figure}
    \centering
    \begin{subfigure}[b]{0.2\textwidth}
        \includegraphics[width=\textwidth,trim={0 6cm 0 6cm},clip]{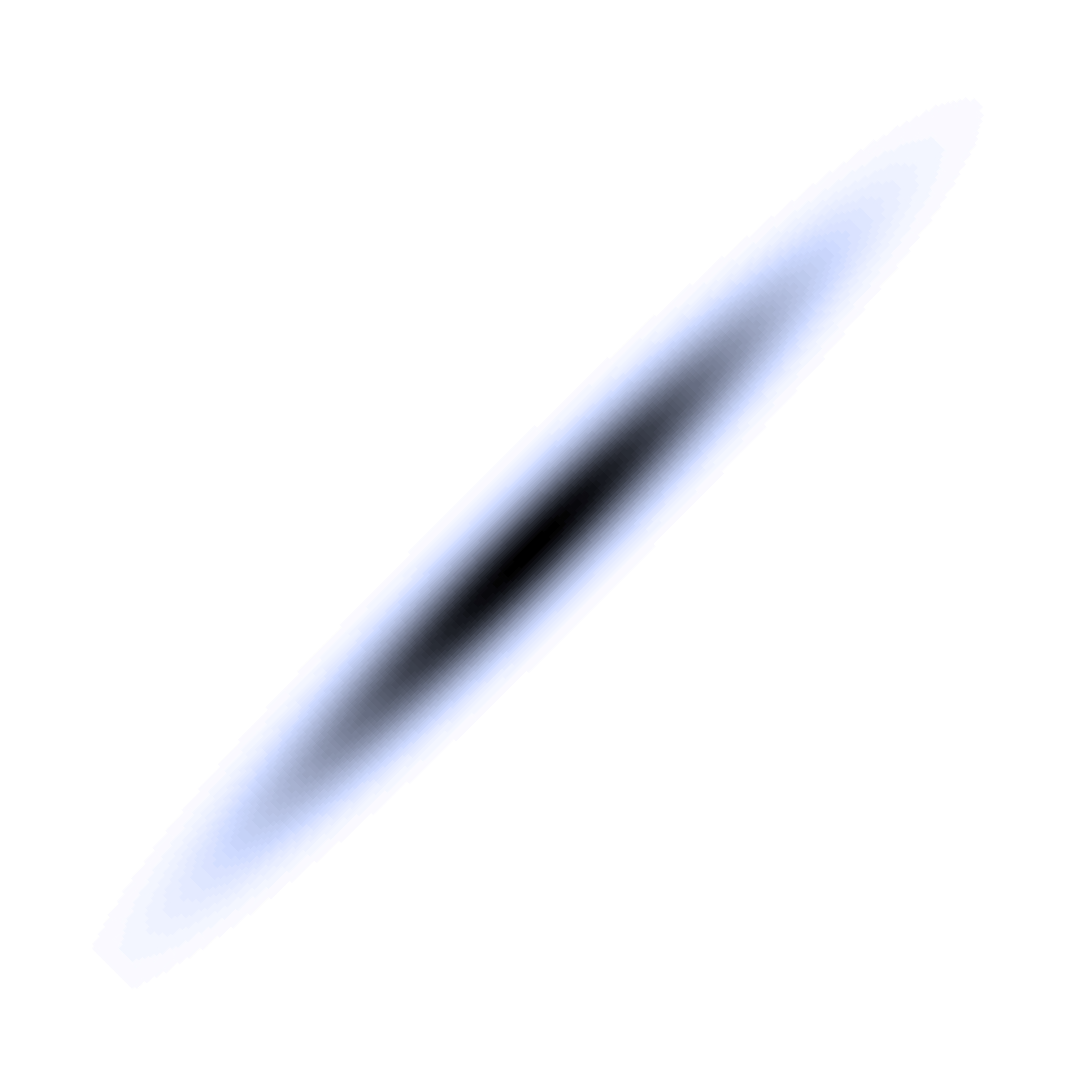}
        \caption{}
        \label{fig:sub_ellipse}
    \end{subfigure}
 \hspace{2cm}
    \begin{subfigure}[b]{0.2\textwidth}
        \includegraphics[width=\textwidth,trim={0 6cm 0 6cm},clip]{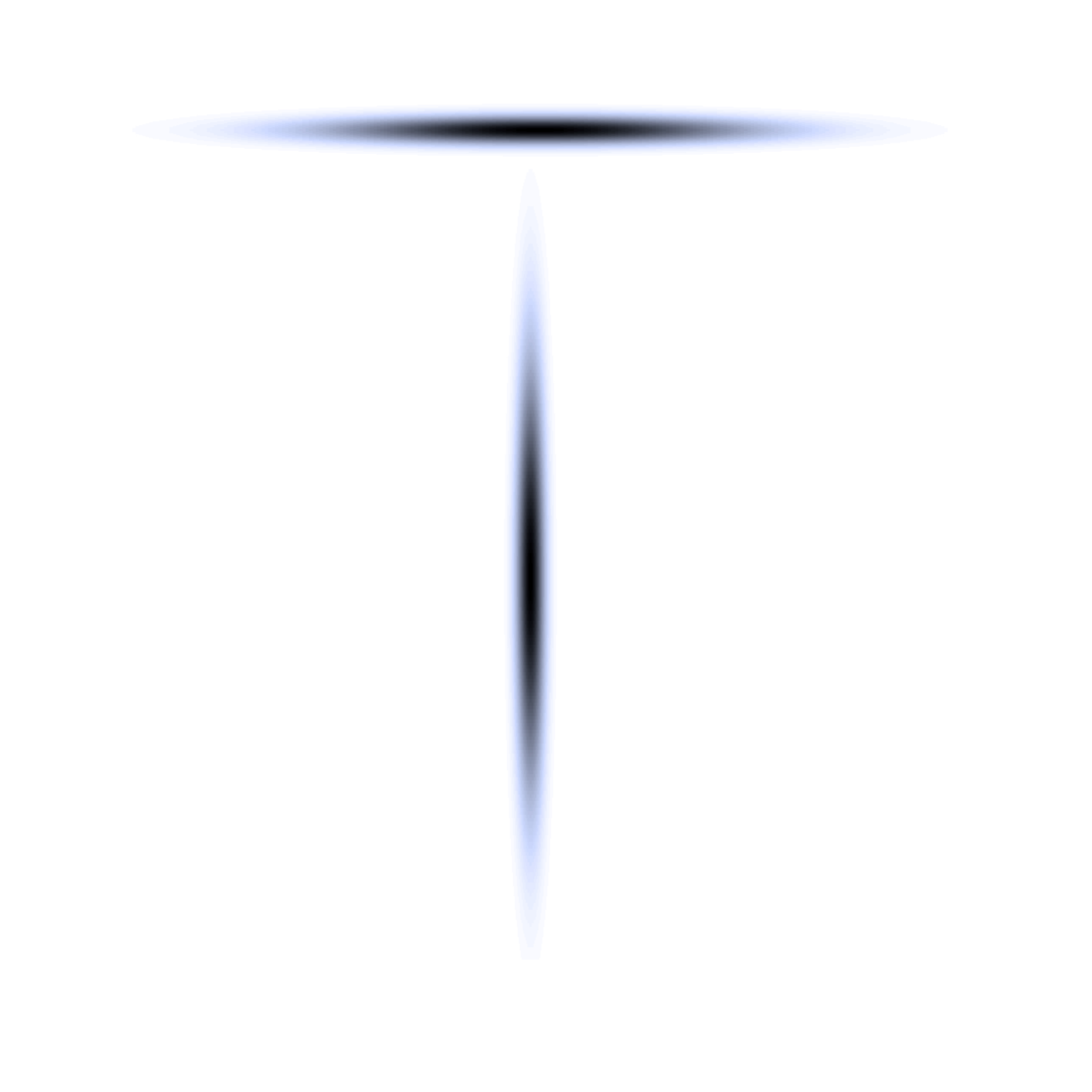}
        \caption{$\,\!$}
        \label{fig:sub_T}
    \end{subfigure}
 \hspace{2cm}
    \begin{subfigure}[b]{0.2\textwidth}
        \includegraphics[width=\textwidth,trim={0 3cm 0 6cm},clip]{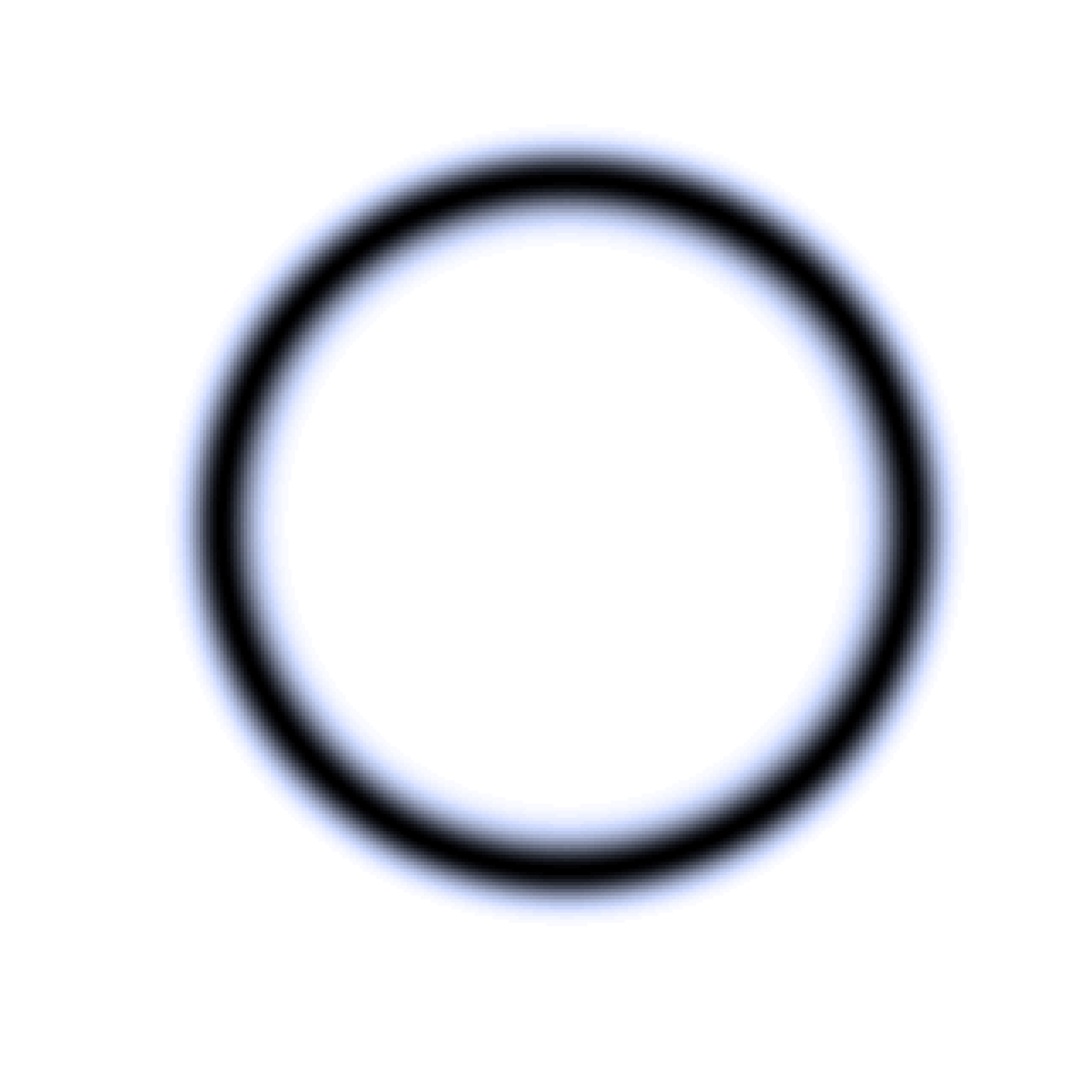}
        \caption{$\!\!$}
        \label{fig:sub_ring}
    \end{subfigure}
    \caption{We plot three examples of posterior distribution functions that might be described as poorly scaled. The distribution (a) has a scaling that can be removed through a linear change of variables, whereas useful scaling information in distributions (b) and (c) depends on the location in space. Proposals can benefit from taking into account local scaling behavior over the global covariance information (the condition number of their covariance matrices in both (b) and (c) are unity). }
    \label{fig::ring}
\end{figure}

More fundamentally,  complex sampling problems will exhibit regions of substantial probability where the Hessian fails to be   positive definite.
A choice that is both less costly and more robust is $S = \Sigma$, where we assume that $\Sigma$ is the covariance matrix of $\pi$ (even when $\pi$ is not Gaussian) and is positive definite.  It can be shown that the iteration in \eqref{eqn::snewton} is again affine invariant.   The resulting scheme, which can be regarded as a simple quasi-Newton type approach, is closely related to adaptive MCMC approaches \cite{RoRo2007,Haario2001}.  On the other hand, because this choice of $S$ does not depend on position, the scheme can be expected to perform poorly on problems for which the scaling behavior is dramatically different in different regions of space (e.g. the Hessian has high condition number and its eigenvectors are strongly position dependent), see Figure \ref{fig::ring}. This observation along with our comments on the method corresponding to $S(x)=-(\nabla^2 \log(\pi(x)) )^{-1}$ suggest a choice of $S$ corresponding to a notion of local covariance.  

While a notion of local covariance will be central to the schemes we eventually introduce, we choose to incorporate that information not through $S$ in \eqref{eqn::general}, but through the skew-symmetric matrix $J$ in that equation.  
In the remainder of this section we discuss how the choices of  $S$ described so far, and the corresponding properties of \eqref{eqn::snewton}, have analogues in choices of $J$ and 
a family of so-called underdamped Langevin schemes that we next introduce.

A popular way to obtain an MCMC scheme with decreased IAT relative to the overdamped scheme in \eqref{eqn::ovdld} is to introduce ``inertia.''  We extend the space by writing our state $x=(q,p)^\text{\tiny T} \in \D \times \R^N \subset \R^{2N}$, with the target distribution $\hat\pi$ written as the product
\begin{equation}\label{phihat}
 \hat\pi(x) = \hat\pi(q,p) = \pi(q) \varphi(p),\qquad \int \hat\pi(q,p) \d p = \pi(q),
\end{equation}
with the distribution of interest $\pi(q)$ recovered as the marginal distribution of the position vector $q$. For the distribution $\varphi(p)$  we will follow common practice and use $\varphi(p) \propto \exp(-\|p\|^2/2).$
With this extension of the space, we can obtain  the standard underdamped form of Langevin dynamics for the choice 
\begin{equation}\label{udlnpre}
 J = \left[ \begin{array}{cc} 0 & -I_N \\ I_N & 0 \end{array}\right], \qquad S = \left[ \begin{array}{cc} 0 & 0 \\ 0 & \gamma I_N \end{array}\right]
\end{equation}
in equation \eqref{eqn::general}, where $I_N$ is the $N\times N$ identity matrix and $\gamma$ is a positive constant \cite{TrMi04}.
Previous work, especially in connection with molecular dynamics \cite{LeMa15}, has examined efficient ways to discretize the Langevin dynamics system while minimizing the error in sampling $\pi(q)$.

To incorporate information such as the Hessian matrix or the covariance matrix (or local covariance matrices) in the underdamped Langevin scheme, we focus on choices of $J$ and $S$ as follows:
\[
J(x) = \left[ \begin{array}{cc} 0 & -B(q) \\ B(q)^\text{\tiny T} & 0 \end{array}\right], \qquad S = \left[ \begin{array}{cc} 0 & 0 \\ 0 & \gamma I_N \end{array}\right],
\]
resulting in the system
\begin{align}
\dot{q} &= B(q) p, \hspace{0.2in} \dot{p} = B(q)^\text{\tiny T} \nabla \log(\pi(q)) + \div( B(q)^\text{\tiny T} )- \gamma p  + \sqrt{2\gamma} \eta(t).
\label{eqn::unewton}
\end{align}
Discretization of the stochastic system may be derived by mimicking the BAOAB scheme \cite{LeMa15}.   Given a stepsize $\h>0$, define $\alpha = \exp(-\gamma\h)$ and approximate the step from $t_n$ to $t_{n+1}=t_n + \h$ by the formulas
\begin{subequations}
\begin{align}
p^{(n+\nicefrac{1}{2})}&=p^{(n)}+\frac{\h}{2} F(q^{(n)})
 ,\qquad \qquad \,
q^{(n+\nicefrac{1}{2})}=q^{(n)} + \frac{\h}{2} B(q^{(n+\nicefrac{1}{2})} )p^{(n+\nicefrac{1}{2})}\label{eqn::discA1}
\\
\hat{p}^{(n+\nicefrac{1}{2})}&=\alpha   p^{(n+\nicefrac{1}{2})}  +\sqrt{1-\alpha^2}\rR^{(n)} + \frac{(\alpha+1)\h}{2}\div(B(q^{(n+\nicefrac{1}{2})})^\text{\tiny T})\label{eqn::discO}
\\
p^{(n+1)}&=\hat{p}^{(n+\nicefrac{1}{2})}+\frac{\h}{2}F(q^{(n+1)}), \qquad q^{(n+1)}=q^{(n+\nicefrac{1}{2})} + \frac{\h}{2} B(q^{(n+\nicefrac{1}{2})})\hat{p}^{(n+\nicefrac{1}{2})}\label{eqn::discB2}
\end{align}
\label{eqn::disc}
\end{subequations}
where $\rR\sim N(0,I_N)$ and $F(q) = B(q)^\text{\tiny T}\nabla\log\pi(q)$.
The choice of matrix $B B^\text{\tiny T}$ introduced in the next section will be a sum of the identity and a small (relative to the dimension $N$) number of rank 1 matrices, alleviating storage demands and reducing the cost of all calculations involving $B$ to linear in $N.$  As described in Appendix \ref{app::metropolization}, schemes of the form in \eqref{eqn::disc} can also be used to generate proposals in a Metropolis--Hastings framework to strictly enforce a condition that, like detailed balance, guarantees that $\pi$ is exactly preserved.  

Suppose that, when applied to sampling the density $\pi_{A,v},$ an underdamped Langevin scheme of the form in \eqref{eqn::disc} generates a sequence $(q^{(n)}, p^{(n)}).$  The scheme  will be referred to as affine invariant if   the transformed sequence $(A q^{(n)}+v, p^{(n)})$ has the same distribution as the sequence generated by the method when applied to sample $\pi.$
As for \eqref{eqn::snewton} one can demonstrate that the choices  $B(q)B^\text{\tiny T}(q) = - (\nabla^2 \log(\pi(q)))^{-1} $  and $B(q)B^\text{\tiny T}(q) = \Sigma,$ yield affine invariant sampling schemes (see Appendix \ref{app::affineinv} for details).

Before proceeding to the important issue of selecting a practically useful choice of $B$, we observe the following important properties of our formulation: (i) the stochastic dynamical system \eqref{eqn::disc} exactly preserves the target distribution and thus, if discretization error is well controlled, Metropolis correction is not likely to be needed, and (ii) the formulation, with  appropriate choice of $B$, is affine invariant, even under discretization (see Appendix \ref{app::affineinv}), a property which ensures the stability of the method under change of coordinates.
By contrast, we emphasize that schemes that modify $S$ (instead of $J$) in \eqref{udlnpre} or that are based on a $q$-dependent normal distribution $\varphi$ in \eqref{phihat} (e.g. within HMC as in \cite{GiCa11}), cannot be made affine invariant in the same sense (Appendix \ref{app::affineinv}).




With the general stochastic quasi-Newton form in \eqref{eqn::disc}  as a template, one may consider many possible choices of $B.$  Just as in optimization, in MCMC the question is not whether one should precondition, but rather how can one precondition in an affordable and effective way.  Unfortunately, practical and effective quasi-Newton approaches for optimization 
do not have direct analogues in the sampling context leaving a substantial gap between un-preconditioned methods and often impractical preconditioning approaches.  In the next section we suggest an alternative strategy to fill this gap: using multiple copies of a simulation to incorporate local scaling information in the $B$ matrix in \eqref{eqn::disc}.

\section{Ensemble quasi-Newton (EQN) schemes \label{LEPSection}}
We next describe an efficient approach to the sampling problem in which  information from an ensemble of walkers provides an estimate of the local covariance matrix.   We consider a system of $L$ walkers (independent copies evolving under the same dynamics) with state $x_i=(q_i,p_i)^\text{\tiny T}$, where subscripts now indicate the walker index. Each walker has position $q_i$ and momentum $p_i$ for $i=1,\cdots,L$, and with $Q=(q_1, q_2, \ldots, q_L)^\text{\tiny T}\in\D^L$ and $P=(p_1,p_2,\ldots,p_L)^\text{\tiny T} \in \R^{NL}$. We seek to sample the product measure $\bar\pi$ whose marginals give copies of the distribution of interest $\pi$:
\[
 \bar{\pi}(Q,P) = \prod_{i=1}^L \hat\pi(q_i,p_i), \qquad \int \bar{\pi}(Q,P) \d P = \prod_{i=1}^L \pi(q_i).
\]
A simple sampling strategy is for each walker to sample  $\bar{\pi}$ by evolving each $x_i$ independently using an equation such as \eqref{eqn::ovdld} or \eqref{udlnpre}.  Such a method scales perfectly in parallel when initial conditions are drawn from the target distribution, but no use is made of the local observed geometry or  inter-walker information.   Alternatively we may use the dynamics \eqref{eqn::unewton} to introduce walker information through the $B(q)$ matrix in order to precondition based on information from the other walkers.  This preconditioning enters into the dynamics but not the distribution $\bar\pi$, ensuring it remains the product of disjoint distributions. Inserting the preconditioning matrix into the distribution in some form can have computational drawbacks associated to the communication of a large amount of information between steps.

Using $L$ walkers, we have the global state $x=(Q,P)$ with $2NL$ total variables, with $B(Q)$ an $NL \times NL$ matrix. We will use $B(Q)=\mathrm{diag}(B_1(Q),B_2(Q),\ldots,B_L(Q))$ with each $B_i(Q)\in\R^{N\times N}$ so that the position and momentum $(q_i,p_i)$ of walker $i$ evolve according to \eqref{eqn::disc} with $B(q)$ replaced by $B_i(Q).$  Note that the divergence and gradient terms in the equation for each walker are are taken with respect to the $q_i$ variable.

With this quasi-Newton framework there are many potential choices for the $B_i$ matrix, with $B_i=I_N$ reducing us to running $L$ independent copies of underdamped Langevin dynamics.  Before exploring the possibilities we remark that, in order to exploit parallelism, we will  divide our $L$ walkers into several groups of equal size.  Walkers in the same group $g(i)$ as walker $i$ will \emph{not} appear in $B_i$ so that the walkers in any single group can be advanced in parallel independently.  
We set
$
Q_{[i]} = \{q_j\,|\,g(j)\neq g(i)\}
$
and let $K$ be the common size of these sets.
For example, if we have 16 cores available we may wish to use  ten groups of $16$ walkers (so $L=160$ and $K=144$). If walker $j$ is designated as belonging to group 1, it evolves under the dynamics given in equation \eqref{eqn::disc} but the set $Q_{[j]}$  only includes walkers in groups $2,\cdots,10$. We may then iterate over the groups of walkers sequentially, moving all the walkers in a particular group in parallel with the others. This compartmentalized parallelism is inspired by a similar approach  in the emcee package \cite{emcee} and provides high parallel efficiency.   


One choice for the preconditioning matrix (not yet the one we employ) is to use the  sample covariance of the ensemble
\begin{equation}\label{globalB}
 B_i(Q) = \sqrt{\cov( Q_{[i]} )}.
\end{equation}
Note that $\div( B_i(Q)^\text{\tiny T} )\equiv0,$ simplifying the Metropolization of the scheme. In order for $B_i(Q)$ to be positive definite, we need at least $N$ linearly independent walker positions
, which at minimum requires that $L>N.$

With the choice of $B_i$ in \eqref{globalB}, the ensemble scheme applied to the density 
\begin{equation} \label{eqn::affineQP}
 \bar\pi_{A,v}(Q,P) = \prod\hat\pi_{A,v}(q_i,p_i)=\prod\hat\pi(Mq_i+v,p),
\end{equation} 
for some invertible matrix $A$ and vector $v,$
generates a sequence of vectors $(q_1^{(n)},\dots, q_L^{(n)}, p_1^{(n)},\dots,p_L^{(n)})$ with the property that the transformed sequence $(A q_1^{(n)} +v, \dots, Aq_L^{(n)}+v, p_1^{(n)},\dots, p_L^{(n)})$ has exactly the same distribution as the sequence generated by the ensemble scheme applied to $\bar \pi$ (see Appendix \ref{app::affineinv}).  Just as choosing $B$ as the square root of global covariance of $\pi$ in \eqref{eqn::unewton} yields an affine invariant scheme, choosing the $B_i$ as the square root of the ensemble covariance yields an affine invariant ensemble scheme.  This affine invariance property suggests that ensemble schemes with $B_i$ chosen as in \eqref{globalB} should perform well when the covariance of $\pi$ has a large condition number.    Note that a related choice in the context of an overdamped formulation and without the measure-correcting $\div(S)$ term (therefore reliant on an accept/reject step) appears in \cite{JGthesis} and is shown to be affine invariant.   An ensemble  version of the HMC scheme using a mass matrix that seems to approximate the expectation of the Hessian matrix  of $\log \pi$ appears in \cite{zhang2011quasi} but we have been unable to verify that the method as described preserves the target density (either exactly or up to a time-discretization error).

Using \eqref{globalB} in our ensemble schemes is problematic for several reasons.  For high dimensional problems, the requirement that $L>N$ may render the memory demands of the methods prohibitive.  This problem can be easily remedied by only approximating and rescaling in the space spanned by the eigenvectors corresponding to the largest eigenvalues of the ensemble covariance matrix.  While such a scheme can be implemented in a reasonably efficient manner, we find that simply blending the sample covariance matrix with the identity via the choice
\begin{equation}\label{Bwnug}
 B_i(Q) = \sqrt{I_N + \mu\,  \cov(Q_{[i]})}, 
\end{equation}
for some fixed parameter $\mu\geq0$ is just as effective and much simpler.  This choice allows $L\leq N$ but destroys affine invariance.  On the other hand as demonstrated in Section \eqref{NTSection}, the method is still capable of dramatically alleviating scaling issues.

Having resolved the rank deficiency issue by moving to the choice of $B_i$ in \eqref{Bwnug}, one difficulty remains.  As described in the previous section, for many problems we might expect that the global covariance of $\pi$ is reasonably well scaled but that the sampling problem is still poorly scaled (the Hessian of $-\log\pi$ has large condition number).  
To address problems of this type, we define a localized covariance matrix that better approximates the Hessian at a point $q_i$ while retaining full rank. We weight samples in the covariance matrix based on their distance (scaled by the global covariance) to a walker's current position, i.e. we use  
\begin{equation} \label{eqn::localB}
 B_i(Q) = \sqrt{ I + \mu\, \wcov(Q_{[i]},\omega_\lambda(Q_{[i]},q_i) ) },
\end{equation}
for parameters $\mu,\,\lambda>0$, where now $\wcov(x,w)$ is a weighted covariance matrix of $K<L$ samples $q \in \R^{K \times N}$ with potentially unnormalized weights $w \in \R_+^K$
\[
 \wcov(q,w) = C\in\R^{N\times N}, \qquad C_{ij} = \sum_{k=1}^K \frac{w_k}{W} ( q_{k,i} - \bar{q}_{i} ) ( q_{k,j} - \bar{q}_j ),\qquad \bar{q}_i = \sum_{k=1}^K \frac{w_k}{W} q_{k,i}
\]
with  $W=\sum_k w_k$ and
\[
 \omega_\lambda(Q,q) = w \in \R_+^{K}, \qquad w_j = \exp\left(-\frac\lambda2 \| Q_j - x\|^2_{\cov(Q)}\right),
\]
where $\|y\|_X^2 = y^\text{\tiny T} X y$.
 Choosing $\lambda=0$ reduces to \eqref{Bwnug}, whereas a large value of $\lambda$ gives more refined estimation of the local scaling properties of the system.
The divergence term in \eqref{eqn::disc} can be computed explicitly by computing partial derivatives of $B_i(q)$, making use of the formula for the derivative of the square root of a matrix:
$\partial_i M(x) = M \Phi(M^{-1} (\partial_i(MM^T) ) M^{-T})$, where $\Phi(M)=\textrm{lower}(M) + \textrm{diag}(M)/2$.
Note that the matrices  $B_i B_i^\text{\tiny T}$ for $B_i$ in \eqref{eqn::localB} are sums of the identity and $L$ rank one matrices so that all manipulations involving $B_i$ can be accomplished in linear cost in the dimension $N.$
In Appendix \ref{app::metropolization} we detail a Metropolis--Hastings test that can be implemented (if needed) to correct any introduced bias.  Because our ensemble scheme preserves $\pi$ exactly when $\deltat$ is small, one can also use the scheme absent of any Metropolis--Hastings test, improving the prospects for it to scale to very high dimension.

\section{Numerical tests \label{NTSection}} 

We consider two numerical experiments to demonstrate the potential improvements that this method offers. A python package with example implementations of the code is available at  \cite{EQN-Code}.

\subsection{Gaussian mixture model}

\begin{figure}[htb]
\begin{center}
\includegraphics[width=.6\textwidth]{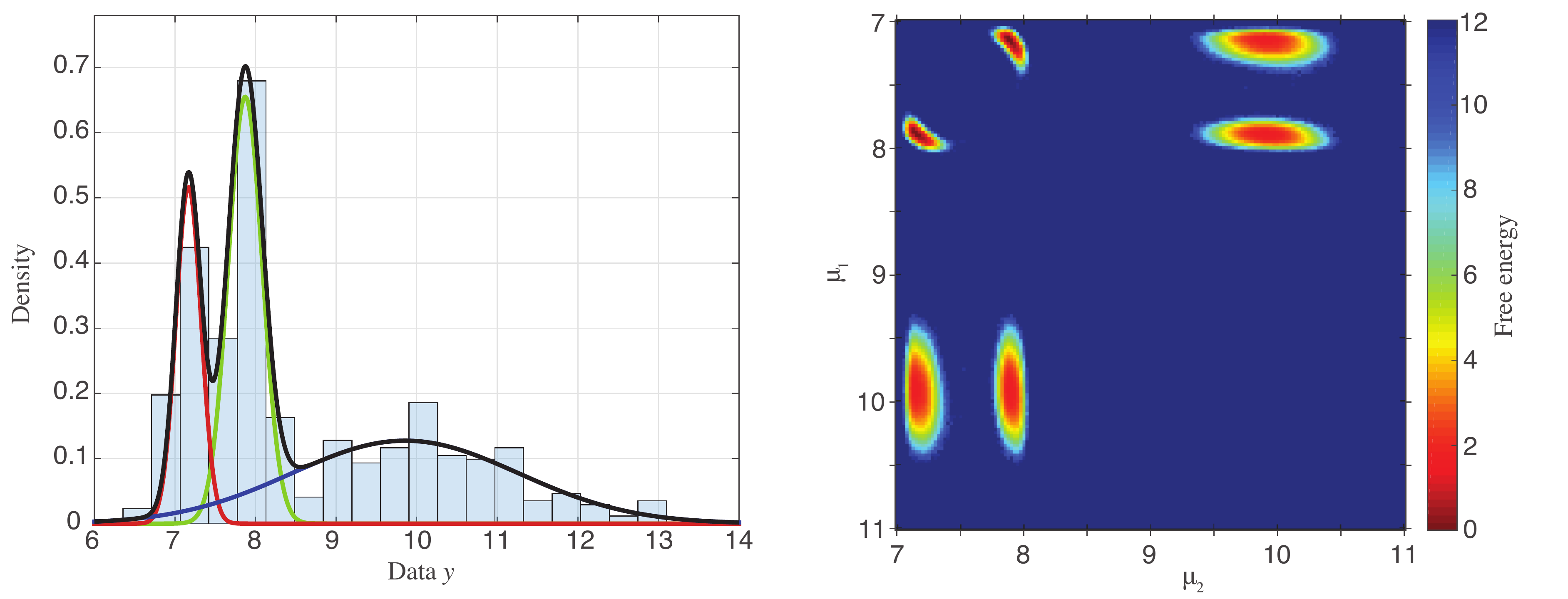}  
\caption{We plot a maximum likelihood state (left) with the three histograms colored as red, green and blue, with their sum in black, along with the original stamp data $y$ as a histogram. The six modes due to label switching  can be seen when looking at the free energy plot (right) in $\mu_1$ and $\mu_2$. }
  \label{fig::gmix}
\end{center}
\end{figure}

We use the model presented in \cite{chopin2012free} focussing on fitting the distribution of a data vector $y$ to univariate mixture model as the sum of three Gaussian distributions. The state vector is described by the means, precisions and weights of the three Gaussian distributions, denoted $\mu_i$, $\lambda_i$ and $z_i$ respectively. Due to the sum of the weights equalling unity, this gives us $N=8$  variables describing the mixture model. We also include a hyperparameter $\beta$ that describes the rate parameter in the prior distribution on the precisions, giving a nine dimensional state. A full description of the problem is available in the Appendix \ref{app::gmix}.
%

We consider the Hidalgo stamps benchmark dataset, studied in \cite{izenman1988philatelic}, as the data $y$ with 485 datapoints. This example is well-suited to the local covariance approach we present above, due to the invariance of the likelihood under a permutation of components (the label-switching problem). Thus the system admits sets of $3!=6$ equivalent modes, see Figure \ref{fig::gmix}, each with a local scaling matrix that has the same eigenvalues with permuted eigenvectors. 

As the walkers may initialize in the neighborhood of different local modes, using a `global' preconditioning strategy would be sub-optimal as the best preconditioning matrix for the current position of a walker depends on which mode is closest to the walker.   Instead, we use the covariance information from proximal walkers as in \eqref{eqn::localB} to determine the optimal scaling. 



We test the EQN scheme against the standard HMC scheme and a Metropolized version of Langevin dynamics. We used $L=64$ walkers for the each scheme and compare the computed integrated autocorrelation times for an ensemble mean of quantities that vary slowly, shown in Table \ref{table::gm}. The autocorrelation times are computed using the ACOR package \cite{ACOR}.

{\small
\begin{table}[tbh]
  \caption{Computed autocorrelation times for slow variables}
  \label{table::gm}
  \centering
  \begin{tabular}{lcccc}
    \toprule  
    Scheme     & $\min(z)$ & $\max(\lambda)$ & $\min(\mu)$ &$\beta$    \\
    \midrule
    HMC & 21495  & 42935 & 27452  & 7148       \\
    Langevin Dynamics     & 6825 & 13279 & 8384 & 4641     \\
    Ensemble Quasi-Newton    & 69 & 83 & 98 & 115  \\
    \bottomrule
  \end{tabular}
\end{table}
}


We consider all three methods as equivalent in cost, as they require the same number of evaluations of $\nabla \log(\pi)$ per step, and scale similarly with the size of the data vector $y$. The EQN scheme is about 100 times more efficient compared to Langevin dynamics, and 350 times more efficient than HMC. We found that removing the divergence term in the EQN scheme had no significant impact on the results.

\subsection{Log Gaussian Cox model}

\begin{figure}[tbh]
\begin{center}
\includegraphics[width=.5\textwidth]{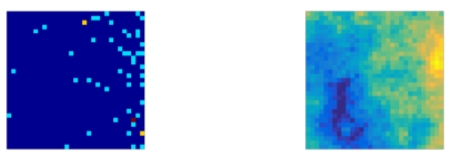} 
\caption{The synthetic observed intensity $Y$ (left) and the true Gaussian field $X$ (right). }
  \label{fig::gcox}
\end{center}
\end{figure}
To illustrate the method in a  high-dimensional setting we compare results for inference in the Log Gaussian Cox point process as in \cite{christensen2005scaling}.  We aim to infer the latent variable field from observation data.
We make use of the RMHMC code template  \cite{RMHMC-Code}.  In the model, we discretize the unit square into a $32\times32$  grid, with the observed intensity in each cell denoted $Y_{i,j}$ and Gaussian field $X_{i,j}$. We use two hyperparameters $\sigma^2$ and $\beta$ to govern the priors, making the dimensionality of the problem $N = 32^2+2=1026$ dimensions. Full details of the model are provided in Appendix \ref{app::gcox}.


As the evaluation of the derivative of the likelihood is significantly cheaper with respect to the latent variables (tests showed computing the hyperparameter's derivatives to be about one hundred times slower), we employ a Gibbs scheme to first sample the latent variables using multiple steps, then perform one iteration for the hyperparameter distribution. 

We generate synthetic test data $Y$, plotted in Figure \ref{fig::gcox}, and compare the HMC and Langevin dynamics schemes to EQN (using 160 walkers) and the RMHMC scheme \cite{GiCa11}. We additionally compare the results using the Langevin dynamics and EQN scheme without Metropolization, as the dynamics themselves sample $\pi$, and the Metropolis test only serves to remove discretization error (which is dominated by the sampling error in this example). 
RMHMC uses Hessian information to obtain scaling data for the distribution. This gives it a significant increase in cost, but improves the rate at which the sampler decorrelates. For the model, the RMHMC scheme requires about 2.2s  per step  whereas the other schemes require about 0.35s per step.

{
\begin{table}[tbh]
  \caption{Maximum autocorrelation times for each variable}
  \label{table::gcox}
  \centering
  \begin{tabular}{lcccc}
    \toprule  
    Scheme     & Latent variables & $\sigma^2$ & $\beta$     & Efficiency \\
    \midrule
    HMC & 800.7  & 1041.6 & 1318.7 & 1.0     \\
    RMHMC     & 2158.9 & 34.0 & 1502.0 & 0.15     \\
    Langevin Dynamics      & 405.1 & 140.6 & 435.3       & 3.5  \\
    Langevin Dynamics (non-Met)     & 81.6 & 20.5 & 136.5       & 11.2  \\
    Ensemble Quasi-Newton       & 71.9 & 49.2 & 239.5       & 5.4  \\
    Ensemble Quasi-Newton   (non-Met)    & 64.4 & 8.8 & 47.8       & 26.8  \\
    \bottomrule
  \end{tabular}
\end{table}
}


The EQN scheme significantly outperforms the other methods, with the slowest motion of the system (the $\beta$ hyperparameter) decorrelating more rapidly than the HMC or Langevin schemes for approximately the same cost. The RMHMC scheme's requires significant extra computation, making it much less efficient than the standard HMC scheme in this example.

%

\section{Conclusion} 

We have presented a sampling algorithm that utilizes information from an ensemble of walkers to make more efficient moves through space, by discretizing a continuous ergodic quasi-Newton dynamics sampling the target distribution $\pi(x)$. The information from the other walkers can be introduced in several ways, and we give two examples using either local or global covariance information. The two forms of the $B_i$ preconditioning matrix are then tested on benchmark test cases, where we see significant improvement compared to standard schemes.

The EQN scheme is cheap to implement, requiring no extra evaluations of $\nabla\log\pi(x)$ compared to schemes like MALA, and needing no higher derivative or memory terms. The scheme is also easily parallelizable, with communication between walkers being required infrequently. The dynamics \eqref{eqn::unewton} are novel in their approach to the introduction of the scaling information and we build on previous work using walkers running in parallel to provide a cheap alternative to Hessian data.

The full capabilities of the EQN method, in the context of complex data science challenges, remain to be explored.  It is likely that more sophisticated choices of $B_i$ are merited for particular types of applications.  The propagation of an ensemble of walkers also suggests natural extensions of the method to sensitivity analysis and to estimation of the sampling error in the MCMC scheme.  Also left to be explored is the estimation of the convergence rate as a function of the number of walkers, which may be possible for simplified model problems.






\small

\bibliographystyle{abbrv}
\bibliography{EQNbib}
\section*{Appendix}
\appendix

\section{Autocorrelation times for poorly conditioned problems} \label{app::iat}

For comparisons of efficiency in MCMC methods, the integrated correlation time (IAT) $\tau$ is often used as a rough measure of the rate of convergence of the chain, giving the approximate time between independent samples. For a function of interest $f(x)$ with mean zero, its IAT $\tau_f$ is
\[
 \tau_f = 1 + \frac{2}{\mathrm{var}[f(x)] } \sum_{n=1}^\infty \mathrm{cov}[f(x^{(n)}),f(x^{(0)})],
\]
where $\E_{x_0}$ denotes the expectation with respect to the initial conditions $x^{(0)}$. We shall consider sampling the distribution $\pi(x)\propto \exp(-x^T M^{-1} x /2)$ for some symmetric positive definite matrix $M$, in order to investigate the efficiency of  the scheme \eqref{eqn::ovdld}
\[
 x^{(n+1)} = x^{(n)} + \deltat\, \Nabla\log(\pi(x^{(n)})) + \sqrt{2 \deltat}\, \rR^{(n)}.
\]
As in  Section VIII of \cite{MultigridMC} we will use the worst case IAT for functions of form $f(x) = v^T x.$

Suppose that $\lambda_\max$ is the largest eigenvalue of $M^{-1}$ with corresponding eigenvector $v_\max$. Then if we choose $x^{(0)}=v_\max$
\[
\E[ x^{(n)} ] = (1-\deltat \lambda_\max)^n v_\max,
\]
and hence we must choose $\deltat$ such that $|1-\deltat \lambda_\max|<1$ to ensure convergence to the invariant mean vector zero. This gives the linear stability condition $\deltat < 2/\lambda_\max$.

We would expect that using a larger value of $\deltat$ would lead to a more rapid decorrelation between samples. We can quantify this by computing the covariance between samples, where
\begin{align*}
 \mathrm{cov}[x^{(n)},x^{(0)}] &= (I-\deltat M^{-1}) \mathrm{cov}[x^{(n-1)},x^{(0)}]\\
 &= (I-\deltat M^{-1})^n M
\end{align*}
from the definition of the update scheme. Similarly for a test function $f(x)=v^Tx$ we have
\[
 \mathrm{cov}[f(x^{(n)}),f(x^{(0)})] = v^T (I-\deltat M^{-1})^n M v,
\]
with integrated autocorrelation time
\[
 \tau_f = 1 + \frac{2}{v^T M v} v^T \left( \sum_{n=1}^\infty (I-\deltat M^{-1})^n  \right) M v.
\]
Assuming the linear stability condition   is satisfied, we can rewrite this expression as
\begin{equation}
 \tau_f = 1+ \frac{2}{\deltat \, v^T M v} v^T  ( M  -\deltat I )  M v = \frac{2v^T M^2 v}{\deltat \, v^T M v} - 1. \label{eqn::tauf}
\end{equation}
Plugging in $v=v_\min$, the eigenvector corresponding to the minimum eigenvalue of $M^{-1}$ (with eigenvalue $\lambda_\min$) we find that for the particular observable $f_*(x)=v_\min^T x$
\[
 \tau_{f_*} = \frac{2  }{\deltat \, \lambda_\min} - 1,
\]
and given the constraint $\deltat < 2/\lambda_\max$ this gives
\[
 \tau_{f_*} > \frac{\lambda_\max}{\lambda_\min}  - 1,
\]
so that even choosing the largest timestep possible, the rate of convergence will be slow whenever $M$ has a wide range of eigenvalues.

\section{Metropolization of discretized scheme} \label{app::metropolization}
In order to improve stability of the scheme, or to correct for numerical bias, we may seek to impose a Metropolis condition on the discretization of the dynamics \eqref{eqn::unewton}. The discretization we use is given in \eqref{eqn::disc}, which we rewrite here for clarity:
\begin{subequations}
\begin{align}
p^{(n+\nicefrac{1}{4})}&=p^{(n)}+\frac{\h}{2} F(q^{(n)})
 ,\label{eqn::disc2B1}\\ 
q^{(n+\nicefrac{1}{2})}&=q^{(n)} + \frac{\h}{2} B(q^{(n+\nicefrac{1}{2})} )p^{(n+\nicefrac{1}{4})}\label{eqn::disc2A1}\\
p^{(n+\nicefrac{2}{4})} &= p^{(n+\nicefrac{1}{4})} + \frac{\h}{2} \div(B(q^{(n+\nicefrac{1}{2})})^\text{\tiny T}) \label{eqn::discxB} \\
\hat{p}^{(n+\nicefrac{2}{4})}&=\alpha p^{(n+\nicefrac{2}{4})}+\sqrt{1-\alpha^2}\rR^{(n)}\label{eqn::disc2O}\\
p^{(n+\nicefrac{3}{4})} &= \hat{p}^{(n+\nicefrac{2}{4})} + \frac{\h}{2} \div(B(q^{(n+\nicefrac{1}{2})})^\text{\tiny T})  \label{eqn::discxB2}\\
q^{(n+1)}&=q^{(n+\nicefrac{1}{2})} + \frac{\h}{2} B(q^{(n+\nicefrac{1}{2})}) {p}^{(n+\nicefrac{3}{4})}\label{eqn::disc2B2}\\
p^{(n+1)}&= {p}^{(n+\nicefrac{3}{4})}+\frac{\h}{2}F(q^{(n+1)}), \label{eqn::discend}\qquad
\end{align}
\label{eqn::disc2}
\end{subequations}
with $\alpha=\exp(-\gamma \delta t)$, $\rR^{(n)}\sim N(0,I)$ and $F(q) = B(q)^\text{\tiny T}\nabla\log\pi(q)$. Note that the step in \eqref{eqn::disc2A1} must be solved implicitly, likely requiring many evaluations of the matrix $B$. However, as this requires no communication between walkers and no evaluations of $\nabla\log\pi(q)$ we consider this a ``cheap'' operation. 

The ratio of transition probabilities necessary for the acceptance rule is 
\[
\frac{T( (q^{(n)},p^{(n)}) \to (q^{(n+1)},p^{(n+1)}) ) }{T( (q^{(n+1)},-p^{(n+1)}) \to (q^{(n)},-p^{(n)}) )} = 
\frac{ f_\alpha( \rR^{(n)} ) }{ f_\alpha(\alpha  \hat{p}^{(n+\nicefrac{2}{4})} - p^{(n+\nicefrac{2}{4})})  } 
\frac{|I+\tfrac12 \h\nabla\otimes B(q^{(n+\nicefrac{1}{2})})  {p}^{(n+\nicefrac{3}{4})}|}{|I-\tfrac12 \h\nabla\otimes B(q^{(n+\nicefrac{1}{2})}) p^{(n+\nicefrac{1}{4})}|},
\]
with derivatives taken with respect to $q$ and where
\[
f_\alpha(x) = \exp\left(-\frac{\|x\|^2}{2(1-\alpha^2)}\right).
\]
In an efficient implementation, the calculation of the derivatives of $B(q)$ (needed for the transition probabilities and the divergence term) is only required once per step, between \eqref{eqn::disc2A1} and \eqref{eqn::discxB}, while the evaluation of the $F(q)$ term is also once per step (after initialization) between lines \eqref{eqn::disc2B2} and \eqref{eqn::discend}. A single step can then be Metropolized using
\begin{itemize}
 \item Set $(q^*,p^*) \leftarrow \textrm{Upd}(q^{(n)},p^{(n)})$
 \item Draw $u \sim U(0,1)$
 \item Compute \[
                U = \min\left(1, \frac{\hat\pi(q^*,p^*)   }{ \hat\pi(q^{(n)},p^{(n)}) } \frac{T( (q^{(n)},p^{(n)}) \to (q^{*},p^{*}) ) }{T( (q^{*},-p^{*}) \to (q^{(n)},-p^{(n)}) )} \right)
               \]
 \item Then if $u < U$ we accept the move and set $ (q^{(n+1)},p^{(n+1)}) \leftarrow (q^*,p^*)$, otherwise we reject the move and flip the sign of the momentum, so set $ (q^{(n+1)},p^{(n+1)}) \leftarrow (q^{(n)},-p^{(n)})$.
\end{itemize}
where the Upd function corresponds to a step of the discretization in \eqref{eqn::disc2}.  Similarly we can perform multiple steps and then accept/reject the trajectory by multiplying the associated transition probabilities. An example implementation in Python is available at \cite{EQN-Code}. For Metropolization of overdamped schemes such as \eqref{eqn::snewton}, see \cite{EricVandNawaf}.

\section{Details of the Gaussian mixture experiment} \label{app::gmix}

In this section we provide the details of the Gaussian mixture experiment for fitting the Hidalgo stamp data $y$ to the density
\[
 \rho(x \,|\,\theta) = \sum_{k=1}^3 z_k N(x\,|\,\mu_k,\,\lambda_k^{-1}),
\]
where $\mu_k$ and $\lambda_k$ are the center and precision of the component Gaussians, respectively, 
with weights $z_k>0$ such that $\sum z_k = 1$.  Let $\theta$ be the parameter/hyperparameter vector  for this model. The target distribution for $\theta$ is $\pi(\theta) \propto p(\theta) \rho(y \,|\,\theta).$

We use the prior distribution $p(\theta)$  such that for $k\in\{1,2,3\}$
\[
 \mu_k \sim N(m,\kappa^{-1}), \qquad \lambda_k \sim \mathrm{Gamma}(\alpha,\beta),\qquad (z_1,z_2,z_3) \sim \mathrm{Dirichlet}_3(1,1,1),
\]
with hyperparameter $\beta\sim\mathrm{Gamma}(g,h)$ and constants $m=\textrm{mean}(y)$, $r=\textrm{range}(y)$, $\kappa=4/r^2$, $\alpha=2$, $g=0.2$, $h=100g/(\alpha r^2)$.

We compare the standard HMC scheme, with a Metropolized Langevin dynamics scheme and the EQN scheme presented in the paper. For each of the schemes, we tweak the stepsize until the acceptance is on average about $75-80\%$. The HMC and Langevin schemes are run by taking 50 steps per single iteration, and using a Metropolis test on the obtained trajectory, while the EQN scheme takes $5$ steps per iteration. The Langevin and EQN scheme used a friction of $\gamma=0.01$. 

All schemes used 64 walkers, which amounts to 64 independent runs for the HMC and Langevin schemes. The EQN run used the localized form of the covariance matrix, with $\mu=100$ and $\lambda=12$ in \eqref{eqn::localB}, however with the weighting kernel only using the Euclidean distance in the $\mu_i$ space, rather than the entirety of $\theta$. We observed that increasing $\mu$ further reduced the acceptance probability significantly.

\section{Details of the log Gaussian Cox experiment} \label{app::gcox}

We run a larger experiment with 1024+2 total parameters to estimate. In the model, we discretize a unit square into a $32\times32$  grid, with the observed intensity in each cell denoted $Y_{i,j}$ and Gaussian field $X_{i,j}$.  

The intensities are assumed to be conditionally independent and Poisson distributed with means $m\Lambda(i,j)=m\exp(X_{i,j})$ for latent intensity process $\Lambda(i,j)$ and $m=1/32^2$. $X=\{X_{i,j}\}$ is a Gaussian process, where $x=\mathrm{vec}(X)$ we have its mean $E(x)=\mu$ and covariance matrix
\[
 \Sigma_{ (i,j),(i',j')} = \sigma^2 \exp( \delta(i,i',j,j') /32\beta ), \quad \delta(i,i',j,j') = \sqrt{ (i-i')^2+(j-j')^2},
\]
with parameters $\mu, \, \sigma^2$ and $\beta$.

Our goal is to sample likelihoods for the latent variables $X$ but also sample values for the hyperparameters $\beta$ and $\sigma^2$, whose priors we assume are exponentially distributed. 

We generate synthetic data $Y$ from a field $X$, created using $\beta=1/33$, $\sigma^2=1.91$ and $\mu=\log(126)-\sigma^2/2$ . We aim to infer $X$ from our synthetic $Y$, along with the hyperparameters used for the model, using the HMC, RMHMC, Langevin dynamics and ensemble quasi-Newton   methods.  We make use of the RMHMC template code for the problem, available at \url{http://www.ucl.ac.uk/statistics/research/rmhmc}.

In order to run multiple highly-resolved simulations, we use a $32\times32$ grid rather than the $64\times64$ grid used in the original version of the problem. This reduces the dimension of the latent variables from 4096 to 1024, which we still consider large enough for a significant test of the samplers' abilities, but this reduction allows us to run for longer and recover more accurate statistical information. However, the change in the model requires us to alter some parameters used in the RMHMC method, for example the timestep, in order to recover good efficiency. The timestep is increased until we reach $75\%$ acceptance (though we do not claim our choice is optimal).

We implement all of the schemes in MATLAB, with each walker running on a single core of identical architecture. For all methods, one iteration uses 50 latent variable steps for each one hyperparameter step. The ergodic property of the Langevin-dynamics based schemes allows us to run without Metropolization, if we are willing to endure some discretization bias that is not removed with further sampling. We argue that in most practical cases, when using a sensible discretization parameter sampling error should always dominate the discretization bias. 

For the ensemble quasi-Newton sampler, we run using 160 walkers using the global covariance formulation for $B_i$ (effectively $\lambda=0$). We use a Gibbs sampling approach to sample the latent variables and hyperparameters, with the schemes using a  $B_i$ for each partition of variables. For the hyperparameters we used $\mu=1$, and for the latent variables we used $\mu=7$. We use five groups of 32 walkers with the walkers within each group running in parallel for ten thousand steps before switching to the next group. The cost of this communication is negligible as it is done so infrequently.

\section{Affine invariance} \label{app::affineinv}
We can extend our notion of affine invariance to the underdamped case, where the system state is $x=(q,p)$, with target distribution $\hat\pi(x)\propto \pi(q) \exp(-\|p\|^2/2)$. We consider affine transformations exclusively of the form $\hat\psi(x)=(\psi(q),p)=(Ax+v,p)$, defining the transformed distribution
\[
 \hat{\pi}_{\hat\psi} \propto \pi_\psi(q)\exp(-\|p\|^2/2), \qquad \pi_\psi(q) \propto \pi(\psi(q)),
\]
for any invertible matrix $A$ and fixed vector $v$. We may apply the discretization \eqref{eqn::disc2} to the density  $\hat{\pi}_\psi$, using some scaling matrix $B_{ {\pi}_\psi}(q)$ that we shall choose later, with $B_{ {\pi}_\psi}(q) = B_\pi(\psi(q))$. The first computation \eqref{eqn::disc2B1} for this density is
\[
p^{(n+\nicefrac{1}{4})}=p^{(n)}+\frac{\h}{2} B_{ {\pi}_\psi}(q)^\text{\tiny T}\nabla\log{\pi}_\psi(q^{(n)}),
\]
which is equivalent to, when writing $y=\psi(q)$,
\begin{equation}
p^{(n+\nicefrac{1}{4})}=p^{(n)}+\frac{\h}{2} B_{ {\pi}}(y^{(n)})^\text{\tiny T} A^\text{\tiny T} \nabla\log{\pi}(y^{(n)}). \label{eqn::ai1}
\end{equation}
Whereas writing $q=A^{-1}(y-v)$ in line \eqref{eqn::disc2A1} and multiplying by $A$ we have
\begin{equation}
y^{(n+\nicefrac{1}{2})} =y^{(n)} + \frac{\h}{2} A B_{ {\pi}}(y^{(n+\nicefrac{1}{2})}) p^{(n+\nicefrac{1}{4})}.\label{eqn::ai2}
\end{equation}
Finally, line \eqref{eqn::discxB} becomes 
\begin{equation}
p^{(n+\nicefrac{2}{4})}  = p^{(n+\nicefrac{1}{4})} + \frac{\h}{2} g(y^{(n+\nicefrac{1}{2})}) \label{eqn::ai3}
\end{equation}
where $g(x)=\div(B_{ {\pi}}(x)^\text{\tiny T}A)$. 

Line \eqref{eqn::disc2O} remains unchanged, given our choice of affine invariant map $\hat\psi$ does not affect the momentum $p$. Suppose now that we choose 
\[
 B_{ {\pi}_\psi}(q) B_{ {\pi}_\psi}(q)^T = A^{-1} C(q) A^{-T},
\]
for some symmetric positive definite matrix $C$, and thus $B_{ {\pi}_\psi}(q) = A^{-1} \sqrt{C(q)}$. For this choice, the discretization steps in \eqref{eqn::ai1}-\eqref{eqn::ai3} become independent of the scale factor $A$, and hence and we obtain a scale-independent sampler when we use \eqref{eqn::disc2}. One such choice is to use the square root of the (constant) covariance matrix $\mathrm{cov}_{\pi_\psi}(q)$ as the $B(q)$ matrix, as  
\begin{equation}
 B_{ {\pi}_\psi}B_{ {\pi}_\psi}^T = \mathrm{cov}_{\pi_\psi}(q) = \mathrm{cov}_\pi(\psi^{-1}(x)) = A^{-1} \mathrm{cov}_{\pi}(x) A^{-T}, \label{eqn::Bcov}
\end{equation}
satisfying the invariance property as the covariance is automatically symmetric positive definite. Alternatively, one may choose $B$ to be the inverse square root of the Hessian matrix of $\log\pi_\psi$, which shares this property (subject to some conditions on the Hessian).

We have shown that the discretization using this choice of $B$ is affine invariant only up to affine transformations in the $q$ variables. However, we may consider dynamics that are invariant under a transformation
\[
 \tilde{\psi}(q,p)  = (A q+v, A^{-T} p),
\]
which acts in both components, where as before $A_i$ is any invertible matrix and $v_i$ is a vector. The dynamics we consider sample the distribution $\tilde{\pi}(q,p)$ ergodically, where
\[
 \tilde{\pi}(q,p) \propto \pi(q) \tilde\varphi (q,p), \qquad  \tilde\varphi (q,p) = \exp(-p^T M^{-1}_{\pi}(q) p/2 - \log| M_{\pi}(q) |/2 ),
\]
so that the marginal distribution of $\tilde{\pi}(q,p)$ in the momenta is $\pi(q)$. The mass matrix $M_\pi^{-1}(q)$ is a symmetric positive definite matrix that may be dependent on position $q$. For the choice of 
\[
J = \left[\begin{array}{cc} 0 & -I\\I & 0 \end{array}\right]\qquad  S = \left[\begin{array}{cc} 0 & 0 \\0 & \gamma I \end{array}\right]\qquad 
\]
in \eqref{eqn::general} we obtain the standard underdamped Langevin dynamics, and  applying it to $\tilde{\pi}(q,p)$ with $\gamma=1$ gives
\[
 \d q = M_{\pi}^{-1}(q) p \dt, \qquad  \d p = \nabla\log\pi(q) \dt + \nabla\log\tilde\varphi (q,p) \dt - p\dt + \sqrt{2 M_{\pi}(q)} \d W_t,
\]
where the gradient $\nabla$ is taken with respect to the position $q$.  Now applying this dynamics  to the transformed distribution instead
\[
 \tilde{\pi}_{\tilde{\psi}}(q,p) \propto \pi_\psi( q ) \tilde\varphi_{ \psi}(q,p) = \pi( A q + v ) \tilde\varphi ( Aq+v,A^{-T}p)
\]
we obtain
\begin{align} \label{eqn::affinemass}
 \d q &= M_{\pi_\psi}^{-1}(q) p \,\dt, \\
 \d p &= \nabla\log\pi_\psi(q) \dt -\tfrac12 \nabla\log|M_{\pi_\psi}(q)| \dt - \tfrac12\nabla(p^T M_{\pi_\psi}^{-1}(q) p)\dt- p\dt + \sqrt{2 M_{\pi_\psi}(q)} \d W_t,
\end{align}
where $\nabla$ denotes gradient with respect to the position coordinates. Changing variables $r=A^{-T}p$ gives
\begin{align*}
 \d q &= M_{\pi_\psi}^{-1}(q) A^T r \,\dt, \\
 \d r &= A^{-T} \nabla\log\pi_\psi(q) \dt - \tfrac12 A^{-T} \nabla\log|M_{\pi_\psi}(q)| \dt - \tfrac12A^{-T} \nabla(r^T A M_{\pi_\psi}^{-1}(q) A^T r)\dt- r\dt + A^{-T} \sqrt{2 M_{\pi_\psi}(q)} \d W_t.
\end{align*} 
Then writing $y=Aq+v$ we have $\pi_\psi(q) \propto \pi(y)$, and if we assume $M$ is chosen such that 
\begin{equation}
M_{\pi_\psi}^{-1}(q) = A^{-1} C(y) A^{-T} \label{eqn::affinereq}
\end{equation}
for some symmetric positive definite matrix $C(q)$, then the dynamics  become 
\begin{align*}
 \d y &= C(y)  r \,\dt, \\
 \d r &=  \nabla\log\pi (y) \dt +  \tfrac12\nabla\log|C(y)| \dt - \tfrac12  \nabla(r^T C(y)  r)\dt- r\dt +  \sqrt{2 C(y)} \d W_t,
\end{align*} 
eliminating the $A$ scaling term in the dynamics and yielding affine invariance with respect to the transformations $\tilde{\psi}$. 

Using the inverse covariance matrix as the mass is one such choice fo the $M$ matrix, as similar to \eqref{eqn::Bcov} we have
\[
 M_{\pi_\psi} = \cov_{\pi_\psi}(q) = \mathrm{cov}_\pi(\psi^{-1}(x)) = A^{-1} \mathrm{cov}_{\pi}(x) A^{-T}
\]
satisfying \eqref{eqn::affinereq}. The inverse Hessian is another such matrix with this property. 

This result applies directly to the RMHMC scheme \cite{GiCa11} which uses a position dependent mass matrix, however it periodically redraws the momentum and sets $\gamma=0$. If the inverse Hessian is used (assuming it remains symmetric positive definite), or another matrix such that \eqref{eqn::affinereq} holds, then the resulting dynamics will be affine invariant under transformations $\tilde{\psi}(q,p)= (A q+v, A^{-T} p)$.

Computationally there is no obvious benefit to including a $p$-scaling term in the affine transformation, as sampling $\pi(q)$ is the ultimate goal of our sampling, and the inclusion of momentum is to increase efficiency. However, the usage of \eqref{eqn::affinemass} may become practically inefficient in the case of ensemble sampling, as the joint distribution will be
\[
\bar{\pi}(Q,P) = \prod_{i=1}^L \pi(q_i) \exp(-p^T_i M_i^{-1}(Q) p /2 - \log|M_i(Q)|/2).
\]
In the target joint distribution the walker positions are no longer independent.  This complicates  Metropolization of the scheme which will now require calculations involving all walkers when any one walker’s position is changed.  Additionally in each walker's dynamics \eqref{eqn::affinemass}, evaluating the $\nabla_{q_i} \bar{\pi}(Q,P)$ term requires computing derivatives of each of the $L$-many $M_i$ matrices, causing a significant amount of additional computational overhead per step.

\section{Noisy estimation of the divergence term} \label{app::noisyestimation}
The divergence of a positive definite matrix appears in many of the schemes we consider above, however the derivative is sometimes prohibitively computationally expensive to obtain, or infeasible to compute analytically. A Metropolization condition can be enforced to recover the correct sampling without this term, but we may instead approximate the  divergence of a matrix $M(x)$ using a random update. Formally, for a small constant $\epsilon>0$ and vector $R\in\R^N$  we have
\[
Z = [M(x+\epsilon R) - M(x)]R = \epsilon \sum_{i=1}^N R_i \partial_i M(x)R + \epsilon^2\sum_{i=1}^N \sum_{j=1}^N R_i R_j \partial_i \partial_j M(x)R + O(\epsilon^3).
\]
If $R\sim N(0,I)$ then taking the expectation of $Z$ yields
\[
\left(\mathrm{E}[Z]\right)_i = \epsilon \sum_{j=1}^N \partial_i M_{i,j}(x) + O(\epsilon^3),
\]
and so $\mathrm{E}[Z] = \epsilon\,\mathrm{div}(M(x)) + O(\epsilon^3)$. This gives a cheap ``noisy'' approximation of the divergence term that will introduce some small bias into the system (depending on the spectrum of $M(x)$).

\end{document}